# Localization of Excitons by Molecular Layer Formation in a Polymer Film

Sudeshna Chattopadhyay and Alokmay Datta
Surface Physics Division, Saha Institute of Nuclear Physics, 1/AF Bidhannagar, Kolkata 700064, India

**Abstract**

Atactic polystyrene of two different molecular weights (560900 g.mol$^{-1}$ and 212400 g.mol$^{-1}$) have been studied as spin-coated films of thickness varying from ~300Å to ~2500Å, i.e., ~2$R_g$ to ~12$R_g$ where $R_g$ is the unperturbed radius of gyration of polystyrene, using x-ray reflectivity and transmission UV-spectroscopy. Electron density profiles along depth of the films show formation of layers parallel to the substrate surface, when the film thickness is below 4$R_g$, with layer spacing ~2$R_g$. The pure electronic singlet $^1A_{1g} \rightarrow {}^1E_{1u}$ indicates exciton interaction throughout film thickness, as evidenced by the lineshape, drop in extinction coefficient and size-dependent blue shift of the corresponding UV peak. Analysis of the total energy of the exciton as a function of film thickness shows that above the thickness of 4$R_g$ the exciton is delocalized over the film thickness through linear, molecular J-aggregates. The separations between these molecules in the two polystyrene samples match with the respective layer spacing obtained from x-ray reflectivity. Transition dipole moments of the molecules in this aggregate are parallel and each dipole is arranged at an angle of ~43° with the axis of the linear chain. Below a film thickness of 4$R_g$, as the molecular layers are formed, the exciton becomes localized within these layers. This localization is clearly indicated by the lineshape becoming comparable to isolated molecular bands, and by both the extinction coefficient and blue shift reaching maximum values and becoming size independent. Localization of excitons due to layer formation has been explained as caused by decrease of cohesion between layers.

# 1. Introduction

Polystyrene (PS) is considered to be the archetype of polymers with pendant chromophores [1-3] where the pendant phenyl groups or benzene rings form excimers involving two rings on the same chain[2]. The excimers are stabilized through configuration interaction between neutral excitation resonance and charge transfer resonance with maximum stabilization achieved when two rings are arranged in a parallel sandwich configuration with a separation of 3 to 3.5Å [3, 4]. Intramolecular excimer formation is shown to follow Hirayama's 'n = 3 rule' [4, 5], which states that a main chain segment of three carbon atoms between two chromophores is required for excimer formation. In a previous report [6] we have demonstrated that adjacent rings in polystyrene are also correlated as physical 'dimers', with stable ground and excited states giving characteristic optical absorption spectra. These 'dimers' are formed as J-aggregates by the excitation transfer interaction between benzene rings and they do not need to involve charge transfer resonance to explain the absorption spectra. The 'dimers' have a considerably stable V-shaped configuration with a separation of about 5 Å between the ring-centres [6]. Our model made no distinction between compositionally adjacent or distant ring-pairs on the chain or between rings in two different chains, as long as the rings are physically adjacent.

The optical spectra, of polystyrene is explained to be due to Frenkel excitons localized in pendant group by polarization fluctuations due to vibrational modes of the polymer molecules, rather than as delocalized energy bands [1, 7-9]. It is to be noted that this conclusion was based largely on the UV absorption spectra of the $^1A_{1g} \rightarrow {}^1B_{2u}$ singlet. But since the vibronic band $^1A_{1g} \rightarrow {}^1B_{2u}$ is electronically forbidden [6], the transition dipole moments are small and exciton delocalization is necessarily over a small length.

On the other hand, the pure electronic singlet $^1A_{1g} \rightarrow {}^1E_{1u}$ with much greater transition dipole moment has been studied only as a part of vacuum ultraviolet spectroscopy of polystyrene [10, 11]. The very intense transition of the solid polystyrene film giving a peak at about 6.3eV (197nm) would normally be ascribed directly to the analog of the $^1A_{1g} \rightarrow {}^1E_{1u}$ benzene transition at about 6.933eV (179nm) [12]. The unexpectedly large red shift and large decrease of extinction coefficient of PS peak due to the $^1A_{1g} \rightarrow {}^1E_{1u}$ transition, on passing from solution to solid phase is already suggested to be caused by exciton-type interaction between the benzene groups of the polymer chains [11]. Thus even in these early studies [10, 11] it was indicated that, due to the much larger transition dipole moments involved in the pure electronic transition, exciton delocalization plays a more dominant role and may be modified by polymer conformation. The larger transition dipole moment of the $^1A_{1g} \rightarrow {}^1E_{1u}$ transition is expected to provide a larger J-aggregate, i.e., a more delocalized exciton, and this increased exciton delocalization length will be more prone to quantum size effects caused by the thinning of the polystyrene film.

The growth of nanotechnology and thin film technology has led to the study of polymer thin films [13], especially the morphological changes that are associated with the confinement of polymer molecules in ultra thin films. It has been found that, with thinning of polystyrene films, the polymer tends to form layers parallel to the supporting substrate [14] though the exact thickness required for commencement of this transition has not been ascertained. It has been suggested that these layers are formed by the so called 'gyration spheres' of the polystyrene molecules, each sphere being the maximum entropy configuration of the molecule, and for polymers with large molecular weights these layers form a stable structure due to chain entanglement [15]. Effect of such layering on the thermal behaviour of polystyrene has been studied

[16]. However, the corresponding effect on spectroscopy, i.e., on the excitonic interaction in polystyrene has not been studied yet.

In this communication we have presented the result of our studies on atactic PS films of large molecular weights (560900 g.mol$^{-1}$ and 212400 g.mol$^{-1}$), spin-coated on fused quartz, as the films are varied in thickness from ~$12R_g$ to ~$2R_g$ where $R_g$ (= $0.272 \times (M_0)^{1/2}$, $M_0$ being the molecular weight of polymer) is the unperturbed radius of gyration of the PS concerned. We have studied the film with x-ray reflectivity to extract the electron density profiles (EDP's) along the depth of the films. We have correlated the results of excitonic interaction in these films as obtained from optical spectroscopy of the $^1A_{1g} \rightarrow {}^1E_{1u}$ band to the EDP's obtained from x-ray reflectivity.

## 2. Experiment

Two series of PS, "PS-5" (Molecular weight = 560900 g.mol$^{-1}$, $R_g$ =204Å) and "PS-2" (Molecular weight = 212400 g.mol$^{-1}$, $R_g$ =125Å) thin films were prepared by spin-coating onto the polished fused quartz substrates (25mm × 25mm × 1mm) from 5.5mg/ml to 15mg/ml toluene solutions, at angular velocities varying from 4.03 to 1.00 krpm using a photo-resist spin-coater (Headway Inc, USA). The quartz substrates were cleaned according to the RCA cleaning procedure using ammonia and hydrogen peroxide solutions. Finally, rinsing the substrates in acetone and ethyl alcohol cleaned organic impurities.

The tacticity of the polystyrene sample was characterized by Transmission IR spectroscopy (Perkin-Elmer Spectrum GX FTIR Spectrometer, USA) with a resolution of 4.0 cm$^{-1}$ at 20°C, ranging from 3600 to 350 cm$^{-1}$. It was confirmed that the polystyrene sample was almost purely atactic, i.e., completely amorphous in its bulk state [6]. Transmission UV spectra of the film of different thickness were recorded (using a GBC Cintra 10e spectrometer, Australia) in the 190nm (6.532eV) to 400nm (3.103eV) scan range at the scan speed of 5nm per minute with 1.5nm slit width. Grazing incidence x-ray reflectivity (XRR) of the films was carried out using an 18kW rotating anode (Enraf Nonius FR591, the Netherlands) x-ray source with Cu K$_{\alpha 1}$ line monochromatized by a Si (111) crystal and a 3-circle Diffractometer [17].

## 3. Analysis Schemes

We have compared the depth density profiles or electron density profiles of the films with their excitonic behaviour as J-aggregates. Here we outline the techniques of analyzing the x-ray data to extract EDP's and of analyzing the UV data to extract the exciton transfer interaction and the *J*-aggregate geometries.

### 3.1. Electron density profiles (EDP's) of the PS films

In x-ray reflectivity measurements, a well collimated monochromatized x-ray beam is made to be incident on the sample surface at grazing angle α (starting from few milliradians) and the reflected intensity is recorded in the plane of incidence at angle β [17]. In specular condition, the incidence angle (α) and scattered angle (β) are equal (α = β = θ, for example) and only the non-zero component of the wave vector (**q**) is given by $q_z$ = (4π/λ)sinθ. Here λ is the wavelength of the x-ray used. When α > α$_c$ the incident x-rays penetrate into the sample and in XRR scans the intensity of the specularly reflected beam is measured as a function of grazing incidence angle. Analyzing the XRR data, the electron density of the film as a function of depth, starting from the air/film interface ($\rho(z)$), viz., the electron density profile can be obtained. To extract the unknown EDP from x-ray reflectivity data without any *a*

*priori* assumption and to detect small variations in EDP, a method has been developed [14]. This method has proved to be particularly useful in detecting layering transitions in disordered media [14, 18]. In this scheme we consider the film to be compared of a number of thin slices of equal thickness and of electron density $\rho_0 + \Delta\rho(z)$ where $\Delta\rho$ varies with slices but is constant in a slice. The specular reflectivity ($R$), given in the Distorted Wave Born Approximation (DWBA) [19, 14] is given by

$$R(q_z) = \left| ir_0(q_z) + \frac{4\pi r_0}{q_z}[a^2(q_z)\Delta\rho(q_z) + b^2(q_z)\Delta\rho^*(q_z)] \right|^2$$

(1)

where $r_0$ is the specular reflectance coefficient of the film with average electron density (AED) $\rho_0$, $a$ and $b$ are the coefficients for the transmitted and reflected amplitudes of the average density film and $\Delta\rho(q_z)$ is the Fourier transform of the variation of electron density, $\Delta\rho(z)$. By selecting an appropriate number of slices and $\rho_0$ of the film we fit equation (1), after convolution with a Gaussian resolution function, to the data, keeping $\Delta\rho$'s as the fit parameters.

### 3.2. Energy and geometry of J-aggregates

Extended electronic excitations are known to exist as Frenkel excitons in molecular J-aggregates [20] and form the so-called 'J-band'. This band-structure indicates that lineshapes for J-aggregates should be different from molecular spectral lineshapes. In our spectral analysis, we consider a chain of $N$ identical, equidistant, polarizable, two-level (separated by energy $\varepsilon$) molecules, at positions $r_1, ..., r_N$, coupled to a quantized multimode electromagnetic field of wave vector $k$ ($k = |k| = \varepsilon/c$) with their transition dipoles of magnitude $\mu$ all oriented parallel to each other. The angle between the transition dipoles and the chain is $\gamma$. The instantaneous intermolecular interaction (transfer integral) between molecules $n$ and $m$, is then given by [21]

$$J_{nm} = \frac{\mu^2}{r_{nm}^3}[1 - 3(\hat{\mu}\cdot\hat{r}_{nm})^2]\cos(kr_{nm}) + \frac{\mu^2 k}{r_{nm}^2}[1 - 3(\hat{\mu}\cdot\hat{r}_{nm})^2]\sin(kr_{nm}) - \frac{\mu^2 k^2}{r_{nm}}[1 - (\hat{\mu}\cdot\hat{r}_{nm})^2]\cos(kr_{nm})$$

(2)

where $r_{nm}$ is the distance between molecules $n$ and $m$ and $\boldsymbol{r}_{nm} = r_{nm}\hat{r}_{nm}$, $\boldsymbol{\mu} = \mu\hat{\mu}$. For molecules small in comparison with the used wavelength ($kL \ll 1$) the expression of $J_{nm}$ becomes [20],

$$J_{nm} = \frac{\mu^2}{r_{nm}^3}(1 - 3\cos^2\gamma)$$

(3)

$\cos\gamma$ being equal to $\hat{\mu}\cdot\hat{r}_{nm}$. We have shown [6] that the benzene 'dimers' in polystyrene have small diagonal and off-diagonal disorder. Assuming the same in the J-aggregates formed by the pure electronic transition, we have $r_n = na$, where $a$ is the average nearest-neighbour distance, and considering only nearest-neighbour interactions (i.e., $J_{nm}=J$ if $|n-m|=1$, $J_{nm}=0$ otherwise), the transition energy of a $j$-exciton level is

$$E_j = \varepsilon + 2J\cos(\frac{\pi j}{N+1})$$

(4)

where, from equation (2), in the general case of molecules of any size,

$$J = 6.275 \times 10^{-4} \times [\frac{\mu^2}{a^3}(1-3\cos^2\gamma)\cos(ka) + \frac{\mu^2 k}{a^2}(1-3\cos^2\gamma)\sin(ka) - \frac{\mu^2 k^2}{a}(1-\cos^2\gamma)]\cos(ka)]$$

(5)

In Equations (4) and (5) $E$, $\varepsilon$, and $J$ are in eV, $\mu$ is in Debye, $a$ is in nm, $k$ is in nm$^{-1}$. We have used $j = 1$ in our analysis since this state contributes the major part of the excitonic spectrum.

## 4. Results
### 4.1. Layering in the polystyrene films due to confinement
Figures 1(a) and 1(b) depict the x-ray reflectivity data (open circle) and fits using equation (1) (continuous line) of PS-5 and PS-2 films, respectively, with thickness varying from 300 to 1150Å. The EDP's of PS-5 and PS-2 films extracted from these fits are shown in Figures 2(a) and 2(b), respectively. These, obviously, furnish the precise thickness of the films. From Figures 2 it can clearly be observed that when the film thickness was decreased from ~840Å for PS-5 and 704Å for PS-2 there is an oscillation in electron density, which indicates a layering in PS along depth (or thickness) [14]. If we assume that the PS film formed entropy-driven 'gyration spheres' in solution [22], the EDP's indicate that below a certain thickness, which corresponds roughly to 4R$_g$ for both polymers, these 'spheres' have formed layers. This behaviour is exactly akin to the molecular layering due to geometrical confinement in simpler liquids [18]. Unlike these liquids, however, polymers have considerable entanglement [23] and substrate [24] effects on the distribution of their segments, which cause deviations from spherical shape of the polystyrene molecules in these layers leading to a deviation in layer spacing from 2R$_g$. The attractive forces between substrate and film also cause a decrease in the layer spacing as the film thickness is decreased as can be seen from Table 1. Absence of such decrease in layer spacing with decrease in thickness for PS-2 indicates that the substrate effect is less pronounced in polystyrene with lower molecular weight. As the surface area of substrate covered per polymer molecule is approximately (½)$nl^2$ where $n$ is the number of monomer segments per polymer molecule and $l$ is the effective length of this segment [25], the observed decrease in substrate effect on polymers with smaller $n$ is understandable.

Above film thickness of 4R$_g$ (top panel of Figures 2(a) and 2(b)) there is a constant electron density throughout the film (except the position near substrate-polymer interface). EDP's of ~1150Å films, shown in the figures, are representatives of this whole class of density profiles. This indicates that above 4R$_g$ thickness of films, the confinement is not enough to cause layering. For films of thickness ≥ 600Å for PS-5 and ≥ 300Å for PS-2 there is a reduction of electron density near the substrate-polymer interface [26] to the value of benzene electron density, but below those values of thickness there is increase of electron density near the substrate to the density of bulk polystyrene. The exact reason for the variation of polymer density at the substrate surface can be known only from the specific nature of the substrate-polystyrene forces and their dependence on polymer film thickness. We shall address these questions in a separate communication. The polymer-air and polymer-quartz interface roughness were found to be ~3.5Å and ~5Å, respectively.

### 4.2. J-aggregates in the polystyrene films
Figure 3(a) shows the shape of the UV spectral peaks due to the pure electronic singlet $^1A_{1g} \rightarrow {}^1E_{1u}$ in two PS-5 films of thickness 600Å (< 4R$_g$) and 2451Å ( > 4R$_g$).

While the thinner film shows a peak shape characteristic of a weakly interacting molecule, the thicker film has a peak that is distinctly asymmetric in shape, starting with a sharper 'edge' and falling off more slowly and smoothly, reminiscent of the band edges seen in systems with large exciton delocalization lengths [27]. In Figure 3(b) the extinction coefficients ($\kappa$) for the singlet transition in different films of PS-2 (open squares) and PS-5 (filled circles) have been plotted against the film thickness. The extinction coefficient is seen to be almost constant for films of thickness < $4R_g$ and starts falling as the thickness is increased beyond this. Decrease of extinction coefficient indicates exciton interaction in PS films [11]. Since, $\kappa_{total} = \kappa_{scattering} + \kappa_{abs}$ [28] where $\kappa_{scattering\ (abs)}$ are the extinction coefficient for scattering (absorption) and, with the increase of excitonic interaction, there is decrease of $\kappa_{scattering}$, the increase in extinction coefficient as the film is made thinner indicates a drop in the correlation length between the polystyrene molecules caused by the excitonic transfer, or in other words, a decrease in the magnitude of intermolecular exciton transfer interactions. This increase in the coefficient and hence the decrease in the interaction magnitude reaches a maximum as the film thickness goes below $4R_g$, i.e., as the molecular layers start forming.

From Figure 3(a) we also notice a blue shift in the singlet peak as the thickness of the film is reduced. This quantum size effect which, perhaps, is observed for the first time in polystyrene indicates clearly the delocalization of excitons over the entire film thickness. The energy of the $^1A_{1g} \rightarrow {}^1E_{1u}$ exciton PS-2 and PS-5 has been plotted respectively as open squares and filled circles in figure 4 against film thickness. It can be seen that the curves follow a similar nature as in Figure 3(b). The exciton energy increases as the film thickness is decreased but below $4R_g$ thickness there is no further increase with film thinning. This indicates that while above $4R_g$ thickness of film the exciton is delocalized over the total film thickness, below film thickness $4R_g$ there is a much smaller length scale along the thickness within the film in which the exciton is localized.

The size effect in polystyrene, observed in Figure 4, was explained by the theory of molecular J-aggregates. It was considered that a linear J-aggregate of $N$ molecules is arranged along film thickness, where $N$ = (film thickness) / (molecular dimension ($a$) along thickness) and $\varepsilon$ is the transition energy of each molecule. The transition dipole moment ($\boldsymbol{\mu}$) of each molecule is aligned parallel to each other and oriented at angle $\gamma$ with the local axis of the aggregate chain. It is to be noted here that by the term 'molecule' we do not a priori mean a polystyrene molecule. Rather, it is an entity carrying the specific transition dipole moment of the molecule we are interested in. The exciton energy curve in Figure 4 above $4R_g$ film thickness was fitted with the expression given in Equation (4) having $\varepsilon$, $a$, $\gamma$ and $\mu$ as the fitting parameters. For the transition considered the magnitude of the wave vector ($k$ = 0.032nm$^{-1}$) is just one order less than the reciprocal of the molecular size along thickness, so the simplified relation given by Equation (3) cannot be used here and the full expression given by Equation (5) had to be used for the resonance transfer interaction (J). The best fit of $E$, as a function of total thickness of the film, from a Levenberg-Marquardt procedure is shown in Figure 4 along with the observed value. In Table 2, the corresponding extracted values of fitting parameters for PS-5, and PS-2 are given. The thickness-invariant portions of Figure 4 could obviously not be fitted to this expression. But, as the energy of these portions is higher in both polymers, the (negative) resonance interaction as given by Equation (4) is either absent or very much reduced below $4R_g$ thickness.

From Tables 1 and 2, it is seen that average molecular size along film thickness or the average nearest neighbour distance, $a$, in films thicker than $4R_g$, as obtained from UV-spectra of J-aggregates match (within error) with the corresponding separations of layered polymer gyration spheres in films of thickness equal or less than $4R_g$, as extracted from XRR. The dimensions obtained from both analyses indicate a size of the same order of $2R_g$ but definitely less than the corresponding values. Polystyrene in the thicker films has been cast from concentrated solutions and the evaporation of the solvent as the films are spread would increase the concentration very rapidly. This would cause deviations from an ideal solution of polystyrene and hence, instead of $R_g$, the molecular dimension of polystyrene in the films would coincide with the Flory radius $R_F = \alpha R_g$ where $\alpha<1$ [25]. Coupled with the substrate-film attractive forces, this deviation non-ideality of the solution would cause a considerable reduction in the molecular dimension along film depth. The separation between the dipoles forming J-aggregates in the films thus can be identified with dimension of polystyrene molecules along depth. Thus our UV-spectroscopic study shows clearly that the polystyrene molecules form linear J-aggregates along the depth of polystyrene film when the film thickness varies between $4R_g$ and $12R_g$. The exciton is delocalized over the J-aggregate leading to quantum size effect [29-31]. The value of $\varepsilon$ then corresponds to the separation of levels in the pure electronic transition for 'isolated' polystyrene molecules forming the J-aggregate. The fact that this values lies between the value for benzene molecules (~6.9 eV) and solid polystyrene (~6.3 eV) is consistent with 'isolated' polystyrene molecules. This energy is also seen (Table 1) to scale with molecular weight, as expected. However, as no independent measurement of such an 'isolated' polystyrene molecule exists to our knowledge, we have to rely on these intuitive arguments about the correctness of our estimates of $\varepsilon$ and $\mu$. The molecular transition dipoles are arranged with their axes making an angle of ~43º with the local axis of the linear chain, similar to other linear molecular J-aggregates [32].

As the film thickness is lowered below $4R_g$, confinement induced layer formation of the molecular gyration 'spheres' takes place along film depth and simultaneously, the exciton becomes localized within these layers. The reduction of density between such molecular layers lowers the cohesive force due to van der Waal's interaction [33] between these layers. This lack of cohesion leads to the discontinuity of the excitonic mode along the thickness of the film, localizing the exciton to individual molecular layers.

**5. Conclusion**

We have shown, through x-ray reflectivity and UV-spectroscopy of polystyrene films of two different molecular weight spin-coated on fused quartz substrates that, the exciton due to $^1A_{1g} \rightarrow {}^1E_{1u}$ transition is delocalized along the entire film thickness for PS films with thickness above $4R_g$. Using the results of the exciton theory for linear J-aggregates for this blue shifted J-band absorption spectrum with decreasing film thickness, some important parameters of these J-aggregates, like transition dipole moment ($\mu$) of each molecule, orientation of $\mu$ with molecular chain axis ($\gamma$), molecular separation ($a$), monomer transition energy ($\varepsilon$), have been determined. We have also shown that, below film thickness of $4R_g$ this exciton becomes localized in the molecular layers formed along film thickness by the gyration spheres.

Layering due to confinement has been found to be a general phenomenon in simple and complex liquids. Hence the results presented here are applicable to molecular J-aggregates confined in films in general and to polymers with pendant chromophores in particular. These results will be useful in modifying optical properties of standard luminescent materials through confinement.

**Table Captions**
1. Structural parameters of the polystyrene samples obtained from x-ray reflectivity studies.
2. Parameters of J-aggregates of the polystyrene samples obtained from UV spectroscopic studies.

Table-1

| Polystyrene Sample | Film thickness (Å) | Layer separation (Å) | |
|---|---|---|---|
| | | Dip to dip | Peak to peak |
| PS-5 | 300 | 201 | 210 |
| | 500 | 212 | 212, 211 |
| | 600 | 217 | 216, 217 |
| | 840 | 261, 245 | 261, 261 |
| PS-2 | 307 | 290 | * |
| | 424 | 289 | * |
| | 624 | 232 | 276, 232 |

* Peak separation indistinct

Table-2

| Fit parameters | Polystyrene sample | |
| --- | --- | --- |
| | PS-5 | PS-2 |
| Energy of 'isolated' molecule, $\varepsilon$ (eV) | 6.489±0.02 | 6.469±0.04 |
| Average molecular separation along thickness $a$ (Å) | 287.32±5.911 | 270.05±12.6 |
| Angle between transition dipole and chain axis $\gamma$ (degree) | 43.81±1 | 43.7±0.7 |
| Molecular transition dipole moment $\mu$ (D) | 4735.27±247 | 3596.29±353 |

**Figure Captions**
1. Reflectivity profile, i.e., plot of x-ray reflectivity as a function of the normal component of momentum transfer ($q_z$) for films of different thickness of (a) polystyrene with molecular weight = 560900g.mol$^{-1}$ (PS-5) and (b) polystyrene with molecular weight = 212400g.mol$^{-1}$ (PS-2). Corresponding film thickness is shown beside each profile. Observed values are in open Circles, while the continuous line are the fits. Profiles have been multiplied by a factor of 100 to shift them upwards for clarity.
2. Electron Density Profiles (EDPs), i.e., plot of average electron density as a function of film depth for films of series (a) PS-5 and (b) PS-2 obtained from corresponding fits of Figures 1(a) and 1(b). The vertical sequence of EDP's follow the vertical sequence of reflectivity profiles from which the EDP's have been obtained.
3. (a) UV absorbance peak corresponding to the pure electronic singlet band $^1A_{1g} \rightarrow {}^1E_{1u}$ for PS-5 films of two thickness (given beside each peak). The dotted vertical line shows the band centre for thicker film. Refer text for details. (b) UV extinction coefficient for the pure electronic singlet band, with respect to the corresponding thinnest films, as a function of films thickness for PS-5 (filled circles) and PS-2 (open squares).
4. Exciton energy of the J-aggregates corresponding to the pure electronic singlet band as a function of film thickness for PS-5 (filled circles) and PS-2 (open squares). The continuous lines, solid for PS-5 and dashed for PS-2 correspond to the fits using the theory of J-aggregates. Refer text for details.

Figures

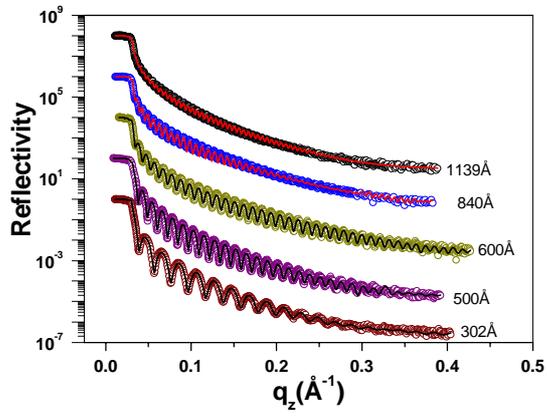
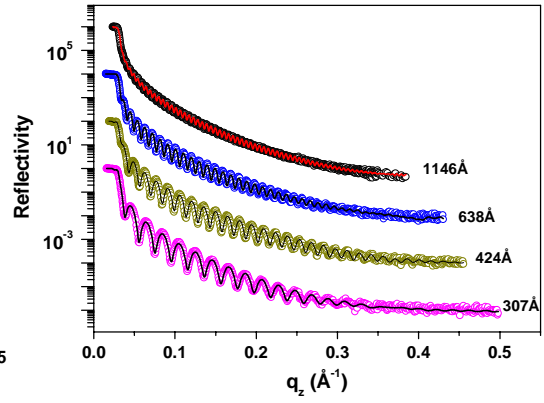

Fig1a

Fig.1b

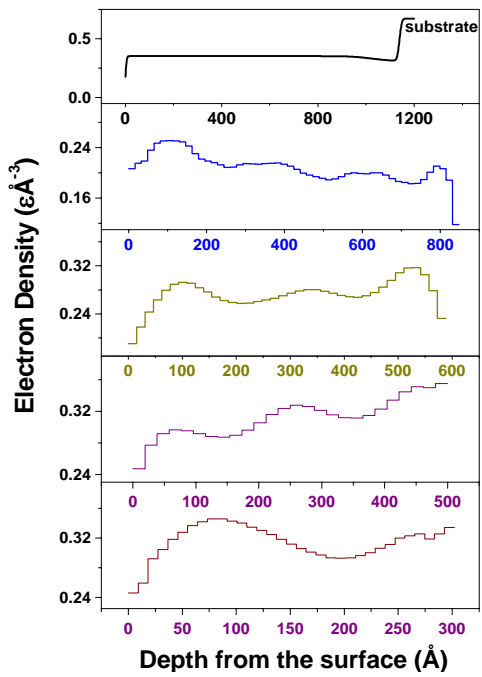
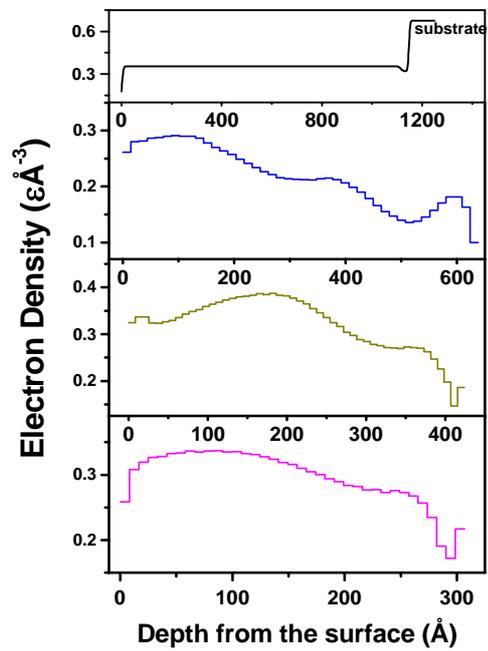

Fig.2a

Fig.2b

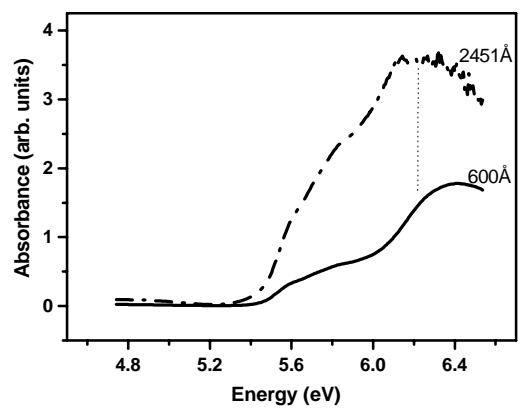 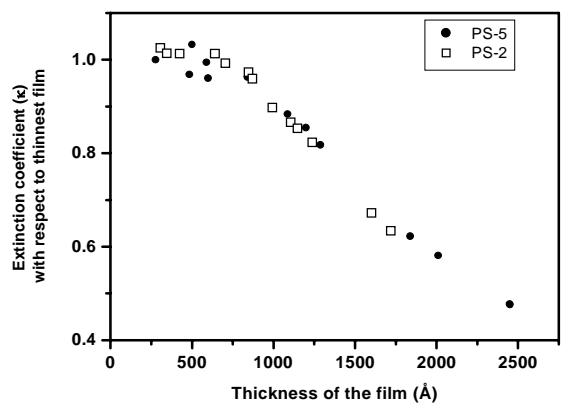

Fig. 3a  Fig. 3b

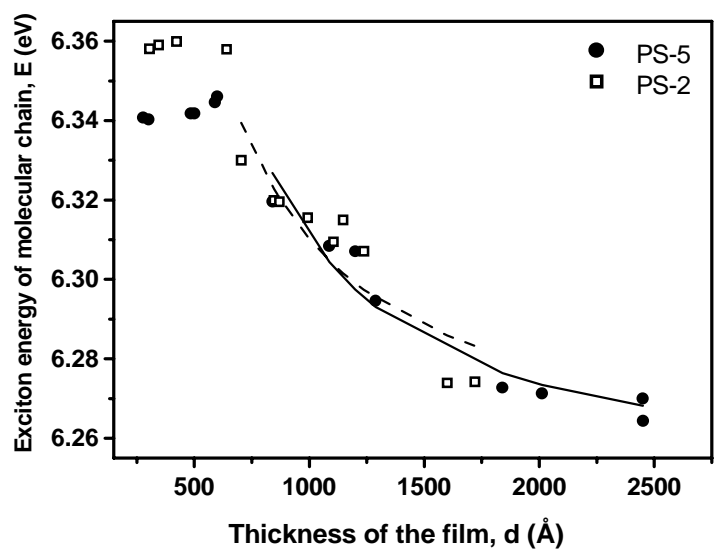

Fig. 4